\documentclass[aip,pop,twocolumn,10pt]{revtex4-1}
\usepackage[pdftex]{graphicx}

\DeclareGraphicsExtensions{.png} 

\usepackage[centertags]{amsmath}
\usepackage{nomencl}
\usepackage{amsfonts}
\usepackage{latexsym}
\usepackage{color}
\definecolor{DarkOlive}{rgb}{0.1047,0.2412,0.0064}
\definecolor{FireBrick}{rgb}{0.5812,0.0074,0.0083}
\definecolor{RoyalBlue}{rgb}{0.0236,0.0894,0.6179}
\definecolor{RoyalGreen}{rgb}{0.0236,0.6179,0.0894}
\definecolor{RoyalRed}{rgb}{0.6179,0.0236,0.0894}
\definecolor{LightBlue}{rgb}{0.8544,0.9511,1.0000}
\definecolor{Black}{rgb}{0.0,0.0,0.0}
\definecolor{FuncColor}{rgb}{1.0,0.0,0.0}
\usepackage{hyperref}
\hypersetup{colorlinks=true,citecolor=blue,linkcolor=blue,urlcolor=black,pdfauthor={Neil P. Oxtoby},pdftitle={Tracking Shocked Dust}}
\usepackage{mathrsfs}
\usepackage[normalem]{ulem} 
\usepackage{subfigure}

\newcommand{\Eq}[1]{Equation~(\ref{#1})}
\newcommand{\sq}[1]{\left[ {#1} \right]}

\newcommand{\ro}[1]{\left( {#1}\right)}

\newcommand{\Ref}[1]{Ref.~\citenum{#1}}
\newcommand{\Fig}[1]{Fig.~\ref{#1}}
\newcommand{\Sec}[1]{Sec.~\ref{#1}}
\newcommand{\Figure}[1]{Figure~\ref{#1}}

\newcommand{\lamD}{\lambda_\mathrm{D}}

\newcommand{\ST}{\rule[-1em]{0pt}{2.5em}} 
\newcommand{\NT}{N_\mathrm{T}}

\begin{document} 
  \title{Tracking shocked dust: state estimation for a complex plasma during a shock wave}
  \author{Neil P. Oxtoby, Jason F. Ralph, C\'eline Durniak and Dmitry Samsonov}
  \affiliation{Department of Electrical Engineering and Electronics, 
  University of Liverpool, Liverpool, \mbox{L69 3GJ}, United Kingdom}
  \date{16 December 2011} %
  \begin{abstract}
    We consider a two-dimensional complex (dusty) plasma crystal excited by an 
    electrostatically-induced shock wave.  
    Dust particle kinematics in such a system are usually determined using particle 
    tracking velocimetry.  
    In this work we present a particle tracking algorithm which determines the dust 
    particle kinematics with significantly higher accuracy than particle tracking 
    velocimetry.  
    The algorithm uses multiple extended Kalman filters to estimate the particle 
    states and an interacting multiple model to assign probabilities to the different 
    filters.  This enables the determination of relevant physical properties of 
    the dust, such as kinetic energy and kinetic temperature, with high precision.  
    We use a Hugoniot shock-jump relation to calculate a pressure-volume diagram from 
    the shocked dust kinematics.  Calculation of the full pressure-volume diagram was 
    possible with our tracking algorithm, but not with particle tracking velocimetry.
  \end{abstract}
  \keywords{complex plasma, dusty plasma, shock, Hugoniot, state estimation, 
  tracking, extended Kalman filter, interacting multiple model}
  \maketitle 

    \section{Introduction}\label{sec:intro}
\begin{figure}[ht]
        {\centering
          {\includegraphics[width=0.99\columnwidth]{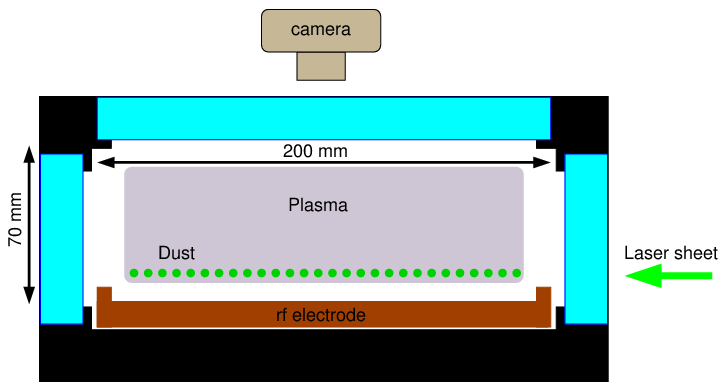}}
          \caption{\label{fig:setup}
            Side-view schematic of the dusty plasma experiment.  
            The dusty plasma is contained wholly within a windowed chamber, with 
            the dust monolayer illuminated by a 2D laser sheet and imaged from above by 
            a digital camera.  
            An offset, below-plane wire (not shown) was used to excite a shock wave in the 
            monolayer as in \Ref{Samsonov2004}.
            }
        }
      \end{figure}

      A complex plasma consists of mesoscopic particles of `dust' suspended within an 
      ionized gas --- a low-density plasma consisting of neutral atoms, ions and 
      electrons.\cite{Merlino2004,Shukla2009:RMP}  
      The dust particles repel each other due to acquiring a net negative charge from 
      collisions with electrons in the plasma (and less frequently, ions).  
      Confined dust can form ordered, crystal-like structures.\cite{Thomas1994:PRL,Melzer1994}  
      In dusty plasma monolayer experiments, single particles of dust are resolvable 
      using a digital camera and illumination by a two-dimensional (2D) laser sheet, 
      making dusty plasma monolayers a useful macromolecular-like system for exploring the 
      microscopic kinematics of condensed-matter systems.  
      The setup shown in \Fig{fig:setup} in \Sec{sec:experiment} can be used in experiments 
      to generate and observe a range of phenomena including Mach cones,\cite{Samsonov1999} 
  	  shock waves,\cite{Samsonov2004}  
  	  tsunamis (steepening effect) and solitons.\cite{Durniak2010:IEEE}  
  	  
  	  Monitoring the dust behavior is a dynamic-state estimation problem.  
  	  Recursive estimation of a dynamic state (position/velocity/acceleration) based on 
  	  remote measurements is often referred to as `tracking', with the tracked object called the 
  	  `target'.\cite{BarShalom,JFRopaedia} 
  	  The most widely-used technique for tracking individual dusty plasma particles based on 
  	  measurements of their positions is known as particle tracking velocimetry (PTV), 
  	  as used in the experiments of Refs.~\onlinecite{Samsonov2000,BoesseAiSR04,Liu2010,Feng2010}, 
  	  for example.  
  	  In PTV, the \emph{average} velocity of a target over one sample period is determined by 
  	  differencing consecutive position measurements.  
      The \emph{instantaneous} velocity of a target can be estimated by including a Bayesian inference 
      step in the recursion.  This prediction step is based on \emph{a priori} knowledge of 
      the target dynamics.  The Kalman filter\cite{Kalman1960,BarShalom,JFRopaedia} is an example 
      of a recursive estimator which combines remote measurements with a predictive step suitable 
      for linear dynamics.  
  	  Nonlinear dynamics can be handled by the extended Kalman filter (EKF),  
  	  which uses a truncated series expansion of the 
  	  nonlinear dynamical equation (usually first- or second-order).  
  	  The Kalman filter is an optimal estimator for linear systems and Gaussian precision, 
  	  in the sense that the minimum mean-square error is obtained.  No such result holds for nonlinear 
  	  systems, but the EKF is widely used because it treats the nonlinearity explicitly and 
  	  typically performs very well when the initial errors and noises are not too large.\cite{BarShalom}  
      A linear Kalman filter has been used to track a single particle in a simulated dust crystal 
      in \Ref{hadziavdic2006}. In that work, the x- and y-dimensions were filtered 
      independently, which limits the accuracy of the prediction. 
      
  	  In this work we present and implement an EKF-based algorithm to track myriad (thousands of) 
  	  dusty plasma particles during the shock wave experiment of \Ref{Samsonov2008:IEEE}, 
  	  and during a computer simulation of a dusty plasma shock wave.  
  	  Central to our algorithm is an interacting multiple model\cite{Blom84,BlomBarShalom88} (IMM) 
  	  tracker which, using a predetermined set of EKFs, automatically switches to the EKF most 
  	  appropriate to the changeable dust kinematics.  
  	  Synthetic data from the computer simulation (described in Section~\ref{sec:dynamics}) 
  	  was used to characterize the algorithm performance in Section~\ref{sec:performance}, showing 
  	  significant improvement over PTV.  
  	  The algorithm was applied to experimental data in Section~\ref{sec:experiment}, 
  	  where a pressure-volume diagram was generated using a Hugoniot shock-jump relation.  
  	  The dusty plasma community is largely unfamiliar with modern target tracking algorithms, 
  	  that originate in aerospace engineering,\cite{BarShalom,JFRopaedia} so we have included 
  	  extensive technical details in an appendix.

    \section{Dust dynamics}\label{sec:dynamics}
      Each charged dust particle in a dusty plasma creates a Coulomb-like potential 
      that is screened by the combination of other particles and the ionized gas.  
      The dust can be treated as point-sources of charge when the 
      radii $R$ are much smaller than the Debye screening length $\lamD$, 
      which in turn is smaller than the inter-particle separation $r$.  
      The effective potential experienced by particle $\jmath$ due to particle $k$ 
      is a screened, repulsive Coulomb interaction of the Debye-H\"uckel/Yukawa 
      form\cite{Konopka2000}:
      \begin{eqnarray}
      	\phi_{\jmath,k}(\vec{r}_{\jmath,k}) &= & 
      	  k_0 Q_\mathrm{d}^2 \frac{e^{-r/\lamD}}{r} , 
      	\label{eq:phi}
      \end{eqnarray}
      where $Q_\mathrm{d}$ is the particle charge, 
      $k_0=1/(4\pi\varepsilon_0)$ is Coulomb's constant, 
      the plasma screening distance is the Debye length $\lamD$, 
      and the particles are separated by $\vec{r}_{\jmath,k} = r \hat{r}_{\jmath,k}$.  
      A hat denotes a unit vector.  
      Taking the negative of the gradient of the two-particle potential yields the 
      effective interaction force between pairs of charged dust particles:
      \begin{equation}
      	\vec{F}_{\jmath,k}(\vec{r}_{\jmath,k}) = 
      	  \frac{k_0 Q_\mathrm{d}^2}{\lamD^2} 
      	  e^{-\tilde{r}} \ro{\tilde{r}^{-2} + \tilde{r}^{-1} } \hat{r}_{\jmath,k} ,
      	\label{eq:force}
      \end{equation}
      where the dimensionless inter-particle separation is defined by $\tilde{r}\equiv r/\lamD$.  
      In monolayer experiments, the dust is confined by an external potential which 
      typically has a shallow parabolic profile (or similar in-plane circular geometry).  
      The dust particles align themselves in an imperfect hexagonal 
      lattice.\cite{Thomas1994:PRL,Melzer1994,Durniak2010:IEEE} 
      Such a dust crystal was the initial condition for the work in this paper: for both 
      experiment and simulation.  
      The particles were subjected to an external force of short duration that induced 
      a shock wave, as in the experiment of \Ref{Samsonov2008:IEEE} and 
      in the simulation Run 7 of \Ref{Durniak2010:IEEE}.  
      The dust was imaged at 1000 frames per second in both the experiment and the computer 
      simulation (where synthetic images were generated).  
      
      The total force on the $\jmath$th particle, $\vec{F}$, is the sum of all two-particle 
      interaction forces $\vec{F}_\jmath = \sum_k \vec{F}_{\jmath,k}$, plus a damping/drag 
      force due to collisions with the plasma, plus the force due to the global confinement 
      potential, and the short-lived external force (see \Ref{Durniak2010:IEEE} for details).

    \section{Algorithm performance} 
      \label{sec:performance}
      The performance of any tracking algorithm is characterized by the errors in the tracks 
      (estimated states).   A computer simulation is necessary for this because the true 
      behaviour of the targets, the `ground truth', is unknown in an experiment.  
      In this section we present errors in the dust particle tracks (position and velocity) 
      using the ground truth from a molecular dynamics simulation of shocked dust.  
      PTV errors were also calculated for comparison.  
      Section~\ref{sub:tracking_algorithm} contains an overview of our tracking algorithm, 
      with extensive details included as an appendix.
      
      The dynamics of the simulated particles were determined using a fifth-order 
      Runge-Kutta numerical integration of the Newtonian equations of motion for 
      position $\vec{q}(t)$ and velocity $\vec{v}(t)$: 
      $d\vec{q}(t)/dt = \vec{v}(t)$, $d\vec{v}(t)/dt = \vec{F}/m$, 
  	  where $m$ is the uniform particle mass (see \Ref{Durniak2010:IEEE} for 
  	  parameter values used in the simulation).  
      
\begin{figure}[htbp]
        \centering
          \includegraphics[width=0.99\columnwidth]{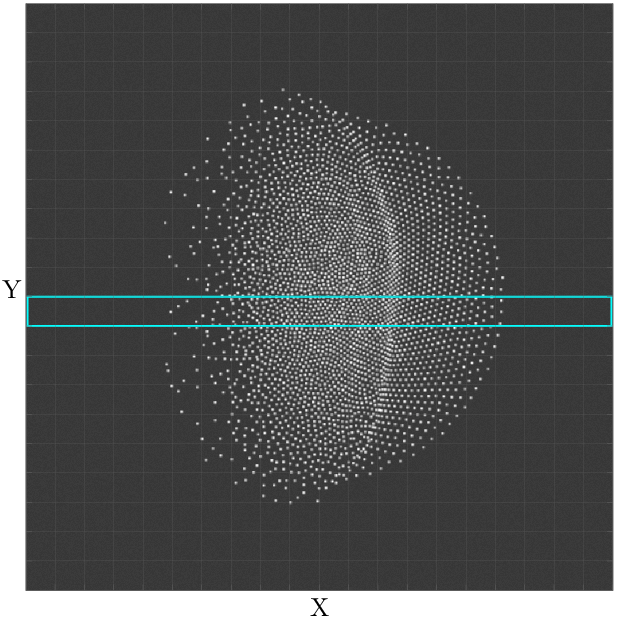}
        \caption{Shock wave simulation.  
          Enhanced grayscale image of a shock wave traversing a dusty plasma.  
          Bulk/average quantities of the dust were calculated for each region of a 
          $20\times20$ grid.  A cross-section `slice' is highlighted.}
        \label{fig:griderize}
      \end{figure}

      We simulated $N=3000$ particles in a weakly-confined 2D dust crystal monolayer, 
      subject to an impulsive external force that induced a shock wave, 
      as illustrated in \Fig{fig:griderize}.  
      The shock wave traversed the captured field of view (the `scene') in one second, with 
      1000 simulated images generated at millisecond intervals ($\Delta{t}=10^{-3}$s).  
      These images were designed to be faithful replicas of those from experiments so that 
      the tracking algorithm was tested with realistic data.  

      \subsection{Tracking algorithm} 
      \label{sub:tracking_algorithm}
        The general procedure for tracking dust (details in Appendix~\ref{app:filter}) starts 
        with particle detection followed by track initialization.  
        Each dust particle typically illuminates a few pixels in an image.  
        The pixel-intensity-weighted centroid of each illuminated 
        region constitutes a particle detection, or measurement.  
        For each subsequent image, track management (also known as track maintenance) 
        associates particle measurements to existing particle tracks, or optionally 
        initiates new tracks from unmatched measurements.  
        Occasionally a particle track will not have a corresponding measurement.  
        Such a missed detection can occur when a dust particle moves out of the 2D laser 
        illumination sheet, or when two particles are close enough to illuminate a single 
        region of pixels.  After a number of consecutive missed detections, a particle 
        track is terminated.  
        The final stage of the tracking procedure is our focus in this paper: state estimation.  
        State estimation uses Bayesian inference to predict the target behavior based on a 
        physical model describing expected target dynamics.  This prediction is combined with 
        measurement in a weighted sum with weightings assigned so as to minimize the expected 
        error in the estimate.  
        For this purpose we used a discrete-time extended Kalman filter (EKF),\cite{BarShalom,JFRopaedia}  
        due to the nonlinear nature of the dust interactions (see Section~\ref{sec:dynamics}).  
        The first-order EKF is a piecewise linearization of the nonlinear dynamics that is used 
        widely in aerospace engineering and works very well in practice.
        
        Kalman filters (extended or otherwise) recursively update the state estimate using a 
        weighted average of predicted and measured values.  The weights are determined by the 
        uncertainties in the prediction and measurement (stored in covariance matrices).  
        These uncertainties are directly influenced by two filter-design parameters known 
        as the process noise and the measurement noise, respectively.  Tuning these design 
        parameters allows one to controllably modify the weighted average between 
        measured and predicted values.  
        
        When the dust is excited by an electrostatic force, the dynamical model used for prediction 
        is augmented by an additional force.  
        In target-tracking nomenclature, this temporary deviation from the 
        original dynamical model is known as a maneuver.  
        Multiple models can be combined in an efficient manner using the interacting multiple model 
        (IMM) approach.\cite{Blom84,BlomBarShalom88,BarShalom}  
        Just as an EKF produces a state estimate by combining a predicted value with a measured 
        value in a weighted sum, the IMM algorithm can be thought of as combining 
        multiple state estimates (from multiple EKFs, for example) in a weighted sum.  
        The weights for each model in the IMM depend on the probabilities of each dynamical model 
        being correct, as determined from the measurement-based likelihoods.  
        Just as the weights in an EKF are directly affected by filter-design parameters 
        (the process noise and measurement noise), the IMM weights are directly affected 
        by selectable parameters known as mode-switching probabilities.  
        As the name suggests, mode-switching probabilities affect how quickly the IMM switches 
        between modes at the onset and at the conclusion of a maneuver.
        
        We used an IMM with three prediction models (EKFs) for the dust dynamics.  
        The first EKF predicted dust behaviour based on the dust interactions.  
        This is known as the base model, or mode, of the IMM.  
        The second EKF model added an external force of fixed magnitude in the positive-X direction, 
        which accounted for the electrostatic excitation force that generated the shock.  
        The third EKF model extended the base model by adding an external force in the negative-X 
        direction, which improved the tracking of particles reflected off the dust cloud bulk.  
        The second and third EKF models are known as maneuvering modes.  
        Technical details of the three EKFs and the IMM are contained in the appendix in 
        sections~\ref{sec:EKF} and~\ref{sec:IMM}, respectively.  
        
        We partitioned the myriad targets into subsets, which were tracked independently 
        across multiple computers using our EKF-based IMM trackers.  
        This partitioning was necessitated by the exponential dependence of computational 
        load on the number of targets per tracker, $\NT$: the multi-target state vector in 
        two dimensions scales as $2\NT$, with the corresponding covariance matrix scaling 
        as the square of this.  
        As discussed in the appendix, this would consume (at double precision) 
        in excess of two gigabytes of memory per time step to track thousands of particles.  
        The subset state estimates were combined offline into a single myriad target 
        state, while ensuring that particles were counted only once each -- there was no 
        unchecked overlap of tracks where a single particle was accidentally tracked by 
        multiple trackers.  
        Such track overlap occur due to particle association errors caused by missed 
        detections or intersecting particle trajectories.  
        Tracker sizes substantially less than 100 are manageable on a desktop computer.  
        Since $\NT$ is a design parameter in myriad target tracking, we considered 
        $\NT=1$, 6 and 12 particle(s) per tracker for comparison.  
        With the goal of minimizing track errors, within limitations imposed by the heavy 
        computational resource requirements of multi-target tracking, we found that 
        target loss was slightly reduced for larger $\NT$, due primarily to the reduced 
        likelihood of track overlap.  Average RMS error levels did not vary significantly 
        with $\NT$.

      \subsection{IMM performance} 
      \label{sub:IMM_performance} %
        The point of the IMM is to detect dust particle maneuvers.  
        Therefore, we measured the IMM performance by comparing the IMM tracker errors with 
        those from a single-mode tracker that did not handle maneuvers: an EKF tracker 
        based on the Yukawa interactions alone.
        
\begin{figure}[ht]
          \centering
          \includegraphics[width=\columnwidth,clip=false,trim=0 18 0 0]{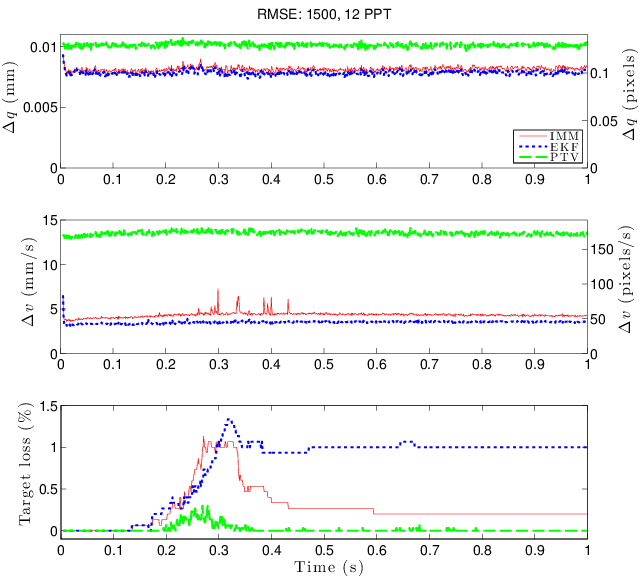}
          \caption{RMS errors in position (top) and velocity (center), and percentage target loss 
          (bottom) as functions of time.  Tracker configuration: $N=1500$ total targets, $\NT=12$ PPT.  
          Compare the IMM tracker (solid red line) 
          with the base mode EKF (dotted blue line) 
          and PTV (dashed green line).  
          }\label{fig:IMM_vs_EKF}
        \end{figure}

        Figure~\ref{fig:IMM_vs_EKF} shows a typical result for all tracker configurations.  
        The bottom plot shows that the IMM trackers were a factor of four better than the 
        EKF trackers (1\% down to 0.25\%) at recovering particles lost during the confusion 
        created by the shock wave formation between $t=0.2$s and $0.4$s (the maneuver).  
        The top and middle plots show that this comes at the cost of increased 
        root-mean-square (RMS) errors for estimated position ($\approx 4\%$ larger error) 
        and velocity ($\approx 20\%$ larger error).  
        This target loss in the simulation was due to the aforementioned track overlap 
        during maneuvers.  
        
      \subsection{Bulk quantity: kinetic energy map}\label{sec:KineticEnergyMap}
        In Figure~\ref{fig:IMM_vs_EKF} we have shown that IMM trackers produce significantly 
        higher-accuracy estimates of the dust particle kinematics than using PTV, particularly 
        for estimating velocity.  In this section we use this higher-accuracy data to estimate 
        a bulk physical quantity: the average kinetic energy
        \begin{equation}
          KE_\mathrm{av} = \frac{1}{2} \overline{m v^2} ~.
          \label{eq:kinetic_energy}
        \end{equation}
        Here $m$ is the particle mass, $v$ is the particle speed, and the overbar indicates 
        the average over multiple particles within each bin in a $20\times20$ grid across the 
        scene (field of view).  The bins are shown in \Fig{fig:griderize}, which contains 
        the simulated image at $t=0.499$s, enhanced for presentation purposes (larger dots 
        and increased brightness).  
        
\begin{figure}[ht]
          {\centering
          \subfigure[~Truth\label{fig:TempMapSnapshotTrue}]{\includegraphics[width=0.48\columnwidth]{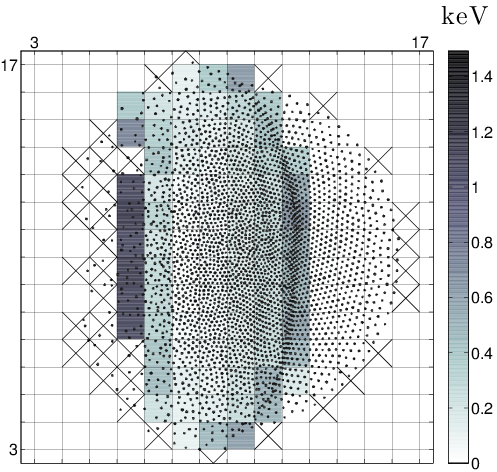}}
          \subfigure[~Estimate\label{fig:TempMapSnapshotEst}]{\includegraphics[width=0.48\columnwidth]{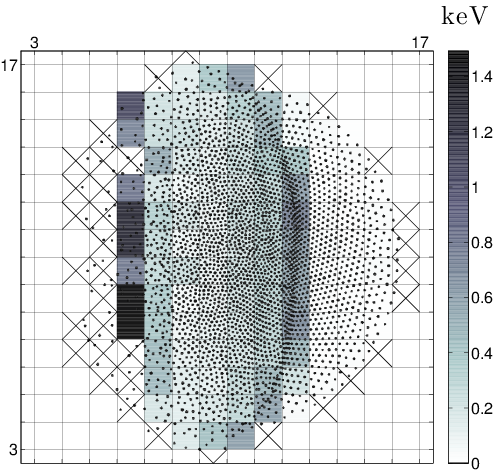}} 
        \caption{Snapshots of the kinetic energy map at $t=0.499$s:  
          \subref{fig:TempMapSnapshotTrue} truth; 
          \subref{fig:TempMapSnapshotEst} IMM-tracker estimate using $N=3000$ with $\NT=1$.  
          The outer three rows/columns of empty bins have been omitted.  
          Particle locations are overlaid for reference.  
          Nonempty bins containing fewer than five particles were ignored and are crossed out.
          \label{fig:TempMapSnapshot}}
          }
        \end{figure}

        A snapshot of the kinetic energy map at $t=0.499$s, estimated using $N=3000$ 
        with $\NT=1$, is shown in \Fig{fig:TempMapSnapshotEst}, with the corresponding true map 
        in \Fig{fig:TempMapSnapshotTrue} showing excellent agreement.  
        Bins containing fewer than five particles are ignored and are crossed out in the figure.  
        Similar results were achieved with $\NT=6$ and 12.  
        
\begin{figure}[!ht]
          {\centering
          \subfigure[~Truth \label{fig:KEmap_slice_true}]{\includegraphics[width=0.49\columnwidth]{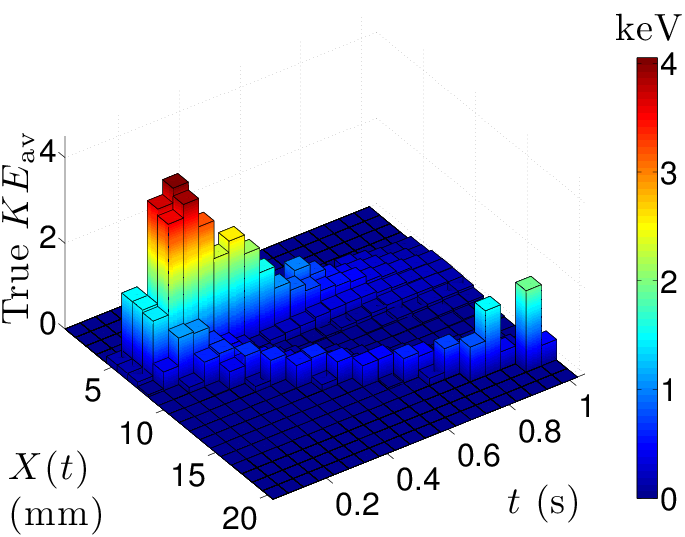}}
          \subfigure[~Estimate\label{fig:KEmap_slice_IMM_3000_1}]{\includegraphics[width=0.49\columnwidth]{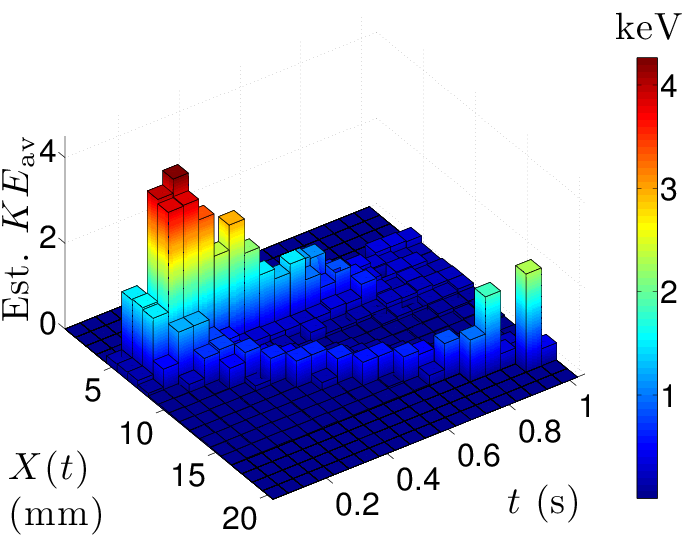}}
          \caption{Kinetic energy map cross-section slice (50ms-spaced snapshots): 
            \subref{fig:KEmap_slice_true} Truth (from the simulation); 
            \subref{fig:KEmap_slice_IMM_3000_1} IMM-tracker estimate using $N=3000$ with $\NT=1$.
            \label{fig:KEmap_slice}
            }
          }
        \end{figure}

        The time-dependence of the accuracy of the estimated kinetic energy map can be 
        visualized by combining selected snapshots of a grid cross-section, or `slice', 
        taken in the direction of shock-wave propagation.  
        We consider the central slice highlighted in \Fig{fig:griderize}.  
        Figure~\ref{fig:KEmap_slice} contains a three-dimensional representation of snapshots 
        of this slice taken every $50$ms, with the height showing the kinetic energy.  
        Figure~\ref{fig:KEmap_slice_true} is the ground truth calculated from the simulation, 
        which compares favorably with the average kinetic energy estimate for in 
        Figure~\ref{fig:KEmap_slice_IMM_3000_1}.  
        The peak of nearly-constant height in the foreground is the shock-front.  
        The highest peak (for small $X$) shows dust particles initially excited by 
        the external force, then later reflected (at lower speed) off the dust cloud.  
        The kinetic energy valley between this excitation/reflection and the shock front 
        is where the dust cloud re-crystallizes in the wake of the shock wave.  

    \section{Experiment} 
    \label{sec:experiment}
      In the experiment reported in \Ref{Samsonov2008:IEEE}, a shock wave was 
      generated by a short sequence of negative voltage pulses along a wire adjacent to 
      the left side of the dusty plasma, and slightly below the plane of the dust cloud.  
      The vertical component of this excitation caused many dust particles to move out of 
      the laser illumination sheet, and hence to disappear, for up to 10 consecutive frames.  
      These missed detections were the primary obstacle to tracking dust particles in the 
      left of the scene.  
      We tracked a total of $N=1860$ particles --- those that were present in all of the first 
      few images.
  	  
\begin{figure}[!ht]
        {\centering
          {\includegraphics[width=0.95\columnwidth]{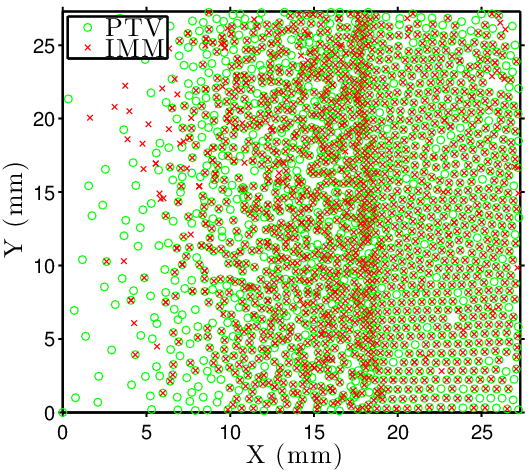}}
          {\includegraphics[width=0.92\columnwidth]{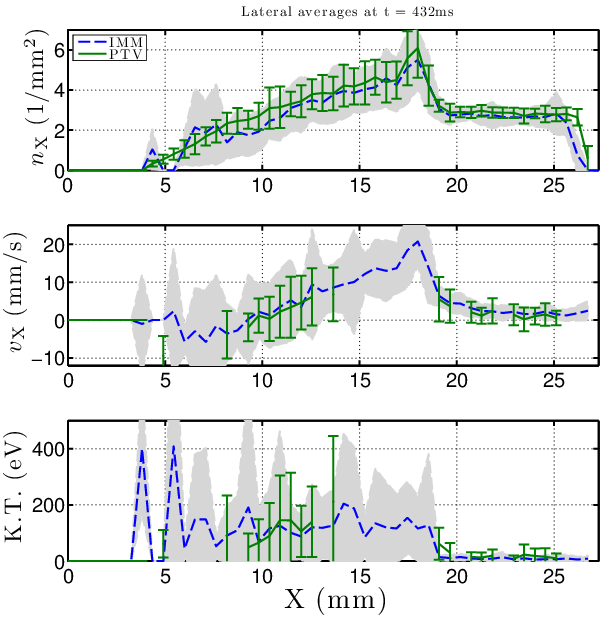}}
          \caption{Tracking shocked dust in an experiment: (top to bottom) snapshot at 
            $t=432$ms of particle positions, particle number density, velocity and 
            kinetic temperature.  
            Standard deviations for each bin are shown as a grey shadow (IMM) 
            and error bars (PTV).  Missing error bars are due to missed detections 
            where PTV is unable to calculate particle velocity.
            \label{fig:experiment_results}
            }
        }
      \end{figure}

      Figure~\ref{fig:experiment_results} shows a snapshot of the experimental results at 
      $t=432$ms.  Shown are the estimated positions
      and average particle number density, velocity, and kinetic temperature in the direction 
      of the shock wave propagation (X).  
      Averages were calculated in each of 50 vertical bins across the scene, with 
      the standard deviation for each bin shown as error bars (PTV) and a shaded region 
      (IMM tracker).  
      The kinetic temperature of particles in each vertical bin was calculated as half 
      the particle mass multiplied by the mean-squared velocity deviation within the bin.  
      Missed detections affected the PTV results severely, with many particle velocities 
      (and consequently kinetic temperature values) missing --- note the large gaps 
      in the PTV plots in \Fig{fig:experiment_results}.  
      Contrast this with the IMM tracker where the problem did not occur.  
      The significant target loss which occurred in the shock formation zone (left of the 
      scene) due to missed detections was not a problem because we were more interested 
      in the shock itself.  
      For a recent discussion of problems and limitations of PTV using high-speed cameras, 
      we refer the reader to \Ref{FengRSI11}.  

\begin{figure}[!ht]
        {\centering
        \subfigure[~$P_2(t)-P_1(t)$\label{fig:PV:IMM:P_time}]{\includegraphics[width=0.65\columnwidth,clip=true,trim=4 0 4 4]{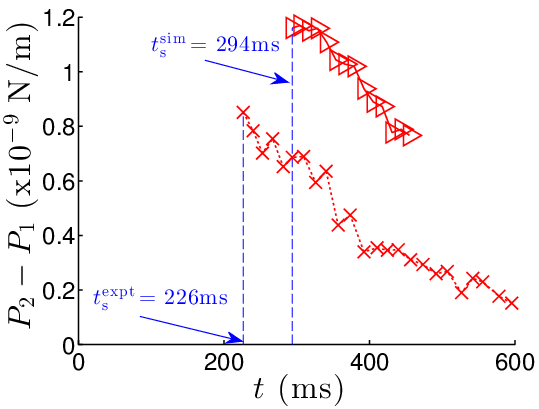}}
        \subfigure[~P-V diagram\label{fig:PV:IMM:P_V}]{\includegraphics[width=0.65\columnwidth,clip=true,trim=4 4 4 0]{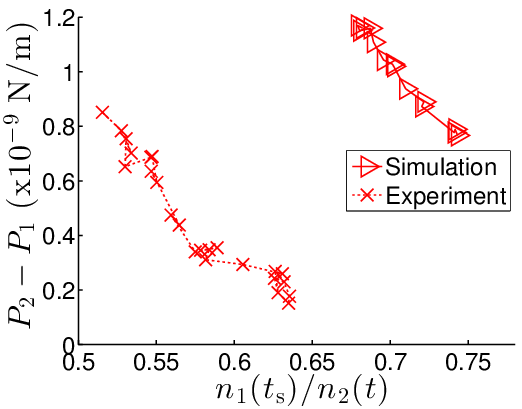}} 
          \caption{Shock-induced pressure jump.  
            Experiment (crosses, dotted line) and simulation (triangles, solid line) 
            showing a steady decrease in shock-induced pressure jump 
            as a function of \subref{fig:PV:IMM:P_time} time, and
            \subref{fig:PV:IMM:P_V} inverse compression.  
            \label{fig:PV}
            }
        }
      \end{figure}

      {The shock-jump (Hugoniot) relations\cite{BondGasDynamics} use conservation laws to 
      relate some physical properties of a fluid upstream (behind) and downstream (ahead) 
      of an ideal shock front.  Specifically, the pressure, density and velocity are 
      related by three equations that also include the shock speed.  
      For known initial conditions downstream, and known shock speed, only two of the upstream 
      properties need to be determined to find the third unknown behind the shock.  
      We consider conservation of momentum across the shock front, which gives the second 
      shock relation\cite{BondGasDynamics}
      \begin{align}
        P_2-P_1 &= m n_{1,2} (U_\mathrm{s} - v_{1,2}) (v_2 - v_1) ~.
        \label{eq:shockJumpP} 
      \end{align}
      Here $P_{1,2}$, $n_{1,2}$ and $v_{1,2}$ are the pressure, number density and particle 
      velocity downstream (1) and upstream (2) from the shock.  
      Our IMM tracker results were used to determine average bin values for $v_{1,2}$ and $n_1$.  
      The shock speed $U_\mathrm{s}=dX_\mathrm{s}(t)/dt$ was determined from the shock front 
      position $X_\mathrm{s}(t)$, which was fitted to the tracker results for peak number 
      density $n_\mathrm{X}(t)$.  
      An ideal shock would travel with constant speed $c_\mathrm{s}$.  
      Real shocks slow down due to damping, which here is a drag force primarily due to 
      ion/neutral collisions with the dust particles.  
      To first order, $U_\mathrm{s}(t) = c_\mathrm{s} - \gamma t$.  
      We determined 
      $c_\mathrm{s,expt} = 42.7 \pm 1.8$mm/s, $\gamma_\mathrm{expt} = 11.1 \pm 1.9$mm/s$^2$ 
      in the experiment, and 
      $c_\mathrm{s,sim} = 53.8 \pm 1.5$mm/s, $\gamma_\mathrm{sim} = 1.7 \pm 1.9$mm/s$^2$ 
      in the simulation.  
      Higher dissipation in the experiment $\gamma_\mathrm{expt}>\gamma_\mathrm{sim}$ 
      suggests either underestimated dissipation in the simulation, or additional unmodelled 
      sources of dissipation.  
      Substituting $m$, $n_1$, $v_1$, $v_2$ and $U_\mathrm{s}$ into equation (\ref{eq:shockJumpP}) 
      yields the shock-induced pressure jump $P_2-P_1$.  
      This is plotted in \Fig{fig:PV:IMM:P_time} as a function of time, and in \Fig{fig:PV:IMM:P_V} 
      as a function of $n_1(t_\mathrm{s})/n_2(t)$ --- the inverse of the shock-induced compression.  
      Here we were restricted to times where the shock was present (after formation at $t_\mathrm{s}$ 
      and before leaving the scene): $t_\mathrm{s}\leq t\lesssim 600$ms.  
      \Figure{fig:PV:IMM:P_V} is essentially a pressure-volume, or P-V, diagram.  
      Experiment (crosses, broken line) and simulation (triangles, solid line) displayed the same 
      qualitative behavior: shock-induced pressure decreased over time and with increasing volume.  
      Pressure decreased over time due to the shock wave losing energy as it traversed the dust 
      cloud.  Pressure was expected to decrease with increasing volume, based on basic 
      thermodynamic arguments (e.g., Boyle's law for an ideal gas).  
      We speculate that the quantitative differences between experiment and simulation were due 
      in part to dust particle uniformity in the simulation versus non-uniformity in the 
      experiment.  For example, we would expect a non-uniform distribution of values for 
      particle mass, charge, etc.~in the experiment.  
      The large gaps in the PTV velocities shown in \Fig{fig:experiment_results} meant that 
      a P-V diagram could not be calculated using PTV. } 

    \section{Conclusion}\label{sec:conclusion}
      We have considered a monolayer dusty plasma crystal disturbed by an 
      electrostatically-induced shock wave, presenting results from a computer simulation 
      and the experiment of \Ref{Samsonov2008:IEEE}.  
      The dust particle kinematics were estimated recursively using an algorithm that 
      combines measurements of position (from images) with predicted behavior based on 
      multiple models for the particle dynamics.  
      {The algorithm provided accurate estimates for the motional states of the dust 
      particles (position, velocity and acceleration) and also selected which of the 
      motion models most accurately reflected the measured data.  }  
      Estimation of dynamic states based on remote measurements is often referred to as 
      target tracking,\cite{BarShalom,JFRopaedia} which here was a nonlinear dynamic 
      estimation problem involving myriad maneuvering objects (up to 3000).  
      Extensive analysis of the results from computer simulations allowed us to quantify 
      the algorithm performance, which was significantly more accurate than using 
      particle tracking velocimetry (a factor of three reduction in the average error).  
      When applied to the experimental data of \Ref{Samsonov2008:IEEE}, the target 
      tracking algorithm demonstrated further superiority to PTV --- 
      the tracking algorithm was robust to missed detections.  Missed detections prevented 
      calculation of particle velocities for PTV.  
      Using Hugoniot shock-jump relations and tracker-estimated kinematics of the dust, 
      the shock-induced pressure jump and compression was calculated for the dust cloud.  
      As expected for a non-ideal shock, pressure decreased as the shock wave lost energy 
      over time.  
      Pressure also decreased as a function of increasing volume (inverse compression).  
      Qualitative agreement was observed between experiment and simulation.
      
      In this paper we have shown that the improvement in precision and reliability of 
      estimating dust kinematics using a target tracking algorithm is important for 
      determining the physics of dusty plasmas.  Our approach can be applied in future 
      experiments to reliably determine quantities such as kinetic energy, kinetic 
      temperature, diffusion coefficients,\cite{Alder1970} and dynamic 
      viscosity.\cite{Gavrikov2005}

\acknowledgments
  This work was supported by the Engineering and Physical Sciences Research Council 
  of the United Kingdom (Grant EP/G007918).  
  N.P.O.~acknowledges use of high-throughput computational resources provided by the 
  eScience team at the University of Liverpool.

\appendix
\section{State Estimation and Tracking}\label{app:filter}
  As overviewed in Section~\ref{sub:tracking_algorithm}, estimation of a dynamic state 
  consists of combining observations (measurements) with predictions.  
  In this appendix we present details of the general procedure used in this paper for 
  tracking \emph{multiple} targets.  It consists of the following steps.  
  Remote measurements of the targets are made, usually at regular intervals.  
  Following track initialization, track maintenance occurs after each measurement.  
  Measurements of the multiple target states (often only positions) are either associated 
  to existing tracks, or used to initiate new tracks.  
  Tracks lacking a measurement (known as a missed detection) are terminated after multiple 
  consecutive missed detections have occurred.  
  State estimation is then implemented and the procedure is repeated for each subsequent 
  measurement.
  
  \subsection{Measurement: image processing}\label{sub:images}
\begin{figure}[!ht]
      {\centering
        \subfigure[~Real image: width $27.3$mm, 1024 pixels\label{fig:experiment_snapshot}]%
          {\includegraphics[width=0.9\columnwidth]{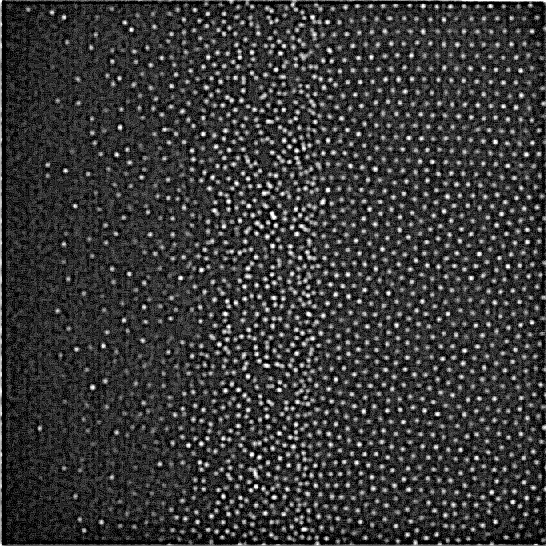}}
        \subfigure[~Synthetic image: cropped width $\approx 27.3$mm, 351 pixels\label{fig:simulation_insets}]%
          {\includegraphics[width=0.9\columnwidth]{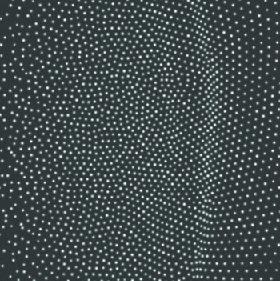}}
        \caption{\label{fig:imageGen2_insets}
          Enhanced images of a dusty plasma shock wave 
          \subref{fig:experiment_snapshot} experiment,\cite{Samsonov2008:IEEE} 
          and \subref{fig:simulation_insets} simulation.  
        }
      }
    \end{figure}

    Dusty plasma observations consist of a time sequence of images taken at (typically) 
    regular intervals separated by $\Delta{t}$.  
    A snapshot at $t=0.353$s from the experiment of \Ref{Samsonov2008:IEEE} is shown in 
    Figure~\ref{fig:experiment_snapshot}.  The image has been enhanced for display 
    purposes (increased particle size and overall image brightness).  
    Similar enhancement was used on the synthetic image in Figure~\ref{fig:simulation_insets}.  
    Our synthetic images emulate the experimental images 
    in Refs.~\onlinecite{Samsonov2004,Feng2007,Samsonov2008:IEEE,RalphSPIE09}: 
    we also use grayscale, 8-bit, tagged image file format (TIFF) images with a $1024$-pixel 
    square field of view.  
    The field of view, or scene, is 27.3mm wide for the experiment, and 80mm wide for the 
    simulation, corresponding to respective spatial resolutions of approximately 
    $26.7\mu$m and $78.1\mu$m per pixel ($37.5$ and $12.8$ pixels per mm).  
    The wider field of view in the simulation enabled the entire dust cloud to be viewed.  
    
    After thresholding the image (setting pixel intensities below some threshold equal to zero), 
    a particle is identified as a contiguous group of bright pixels.  
    Using the moment method, the intensity-weighted centre-of-mass of this group of 
    pixels gives the measured/detected position of a particle\cite{Ivanov2007,Feng2007}:
    \begin{equation}
      \vec{z}_\mathrm{p} = \sum_{\jmath=1}^{N_\mathrm{p}} \frac{I_\jmath}{I_\Sigma} \vec{x}_\jmath ,
      \label{eq:MM_center_of_mass}
    \end{equation}
    where the intensity of each bright pixel is $I_\jmath$, 
    and $\vec{x}_\jmath$ is the $2\times1$ vector containing the x- and y-pixel number 
    for the $\jmath^\mathrm{th}$ pixel in the group containing $N_\mathrm{p}$ contiguous bright pixels.  
    Here the total intensity of the group of bright pixels is 
    $I_\Sigma \equiv \sum_{\jmath=1}^{N_\mathrm{p}} I_\jmath$.  
    For particles illuminating more than a few pixels, the moment method yields a typical 
    measurement precision of around one tenth of a pixel in each direction.  
    More advanced techniques can improve this precision, such as those considered 
    in Refs.~\onlinecite{Ivanov2007,Feng2007}, at the cost of heavier computational load.  
    {However, the potential improvement is limited if measurements are not the dominant source 
    of inaccuracies.}  
    
  \subsection{Track maintenance}\label{sec:measTrack}
    Maintaining continuous tracks involves a decision process known as 
    `measurement-to-track association'.  
    Each measurement is associated to a track, dismissed as a false alarm, or saved 
    as a possible new track.  
    Missed detections must also be handled, since object detection can never be perfect.  
    For example, dust particles may move out of the two-dimensional laser sheet and no longer 
    be illuminated.  We deleted tracks after nine consecutive missed detections.  
    Measurement-to-track association is performed online as part of the tracking algorithm.
    
    Track maintenance is complicated when partitioning myriad targets.  
    Unless using centralized, online data fusion for all trackers (i.e., a parallel tracking 
    algorithm), track `overlap' may occasionally occur.  This is where a single 
    particle is tracked by multiple trackers/computers.  Track overlap often results when 
    particle trajectories intersect.  
    Here we delete such duplicate tracks as part of our offline data fusion process.
    
  \subsection{The extended Kalman filter}\label{sec:EKF}
    We process the measurements and estimate the dust particle states 
    (two-dimensional position $q$, velocity $v$ and acceleration $a$) with a discrete-time 
    extended Kalman filter (EKF).\cite{BarShalom,JFRopaedia}  
    The state of each tracked particle at time step $k=0,1,2,\ldots$ is contained in 
    a six-element vector: 
    $\hat{x}(k)=(q_\mathrm{x}, v_\mathrm{x}, a_\mathrm{x}, q_\mathrm{y}, v_\mathrm{y}, a_\mathrm{y})^T$.  
    The uncertainty in this state is estimated and stored in a corresponding $6\times6$ 
    covariance matrix $S_{\hat{x}}(k)$.  
    Each step in the filter recursion follows measure-update-predict, with the state updated 
    after a measurement as in the standard Kalman filter: 
    \begin{equation}\label{Update}
      \hat{x}(k_{+}) = \hat{x}(k) + K(k)\cdot \sq{z(k) - z(k|k-1)} .
    \end{equation}
    Here the innovation $\sq{z(k) - z(k|k-1)}$ is weighted by the Kalman gain 
    $K(k) = S_{\hat{x}}(k)\cdot H^T\cdot [H\cdot S_{\hat{x}}(k)\cdot H^T+R(k)]^{-1}$.  
    The innovation is the difference between the actual measurement of position $z(k)$ and 
    the expected measurement 
    $z(k|k-1)=H\cdot \hat{x}(k)=(q_\mathrm{x} (k), q_\mathrm{y} (k) )^T$.  
    Here the measurement matrix is 
    \begin{equation}\label{MeasurementMatrix}
      H=\left(\begin{array}{cccccc}
      	1 & 0 & 0 & 0 & 0 & 0 \\
      	0 & 0 & 0 & 1 & 0 & 0 
      \end{array}\right) .
    \end{equation}
    The Kalman gain is a matrix version of the ratio of track uncertainty to total uncertainty 
    (track plus measurement), with the measurement uncertainty given by $R(k)$, the expected 
    covariance matrix for the measurement.  
    Note that matrix multiplication applies throughout --- the dots are to guide the eye.
    
    The updated state $\hat{x}(k_+)$ is now predicted forward to the next time step using a 
    discretization of the governing dynamical process equations given in \Sec{sec:dynamics}:
    \begin{align}\label{nonlinStateTrans}
      {\hat{x}}(k+1) &= f(\hat{x}(k_{+}))\nonumber \\
      &=\left(
      \begin{array}{c}
        \ST q_\mathrm{x} (k_{+}) +(\Delta t)v_\mathrm{x}(k_{+})
        + (\Delta t)^2 F^{\jmath}_{\mathrm{x}}(\hat{x}(k_{+}))/2m\\ 
        \ST v_\mathrm{x}(k_{+}) 
        + (\Delta t)F^{\jmath}_{\mathrm{x}}(\hat{x}(k_{+}))/m\\  
        \ST F^{\jmath}_{\mathrm{x}}(\hat{x}(k_{+}))/m \\ 
        \ST q_\mathrm{y} (k_{+}) + (\Delta t)v_\mathrm{y}(k_{+}) + (\Delta t)^2 F^{\jmath}_{\mathrm{y}}(\hat{x}(k_{+}))/2m\\ 
        \ST v_\mathrm{y}(k_{+}) + (\Delta t)F^{\jmath}_{\mathrm{y}}(\hat{x}(k_{+}))/m\\ 
        \ST F^{\jmath}_{\mathrm{y}}(\hat{x}(k_{+}))/m \\ 
      \end{array}
      \right) 
    \end{align}
    where $\Delta t$ is the time step.  
    Here we have ignored higher-order terms from the Taylor expansion of the full dynamical 
    model equation, which includes the (Gaussian random) process noise $\nu(k)$: 
    $\hat{x}(k+1) = f(\hat{x}(k_+)) + \nu(k)$.  
    The process noise assigns a selectable Gaussian uncertainty to the predicted state 
    $f(\hat{x})$, which increases the state covariance by $Q(k)$, the process noise covariance (matrix).  
    For the first-order EKF the covariance in the predicted state becomes
    \begin{equation}\label{nonlinErrorTransition}
      S_{\hat{x}}(k+1) = f^\prime(\hat{x})\cdot S_{\hat{x}}(k_{+})\cdot f^\prime(\hat{x})^T + Q(k) ,
    \end{equation}
    where the first-order term from the aforementioned Taylor expansion involves the Jacobian matrix 
    \begin{equation}\label{nonlinErrorTransition2}
      f^\prime(\hat{x}) = \left.\frac{\partial f}{\partial \hat{x}}\right |_{\hat{x} = \hat{x}(k_{+})} .
    \end{equation}
    In the standard Kalman filter, the Jacobian is replaced by the prediction matrix that 
    represents the linear dynamics.  As there is no truncated series expansion in the standard 
    Kalman filter, the covariance is a true representation of the prediction error for linear 
    dynamics with Gaussian uncertainty.  In the EKF, the covariance in the prediction is an 
    approximate representation of the prediction error.  
    The first-order EKF is a piecewise linearization of the nonlinear particle dynamics.  
    This linear approximation should be accurate while the errors in the state estimates 
    are small compared to the nonlinear nature of the underlying dynamical 
    function.\cite{JFRopaedia}  
    For our purposes, when the dust particles are in a crystalline state, this approximation 
    is excellent.  In more dynamic moments (during the shock wave, for example), the approximation 
    begins to break down and so the tracker performance is expected to suffer, but not to fail.  
    This is because the process noise in $Q(k)$ affords us the ability to assign a weight to 
    our confidence in the accuracy of the linearized prediction.  
    Similarly, the measurement noise in $R(k)$ reflects our confidence in the measurement 
    accuracy.  These values can be controllably varied to improve the filter accuracy (see 
    \Ref{Oxtoby2010:SPIE} for an example of such filter `tuning').
    
    Larger linearization errors can be dealt with by assigning a lower weight to the prediction, 
    relative to the measurement.  Alternatively, more advanced techniques can be used.  
    These include unscented filters\cite{Julier2004} and particle 
    filters.\cite{Mahler2003:IEEE,ParticleFilterTutorial} 
    
    When tracking multiple targets with a single tracker, the single-particle state vectors 
    are stacked to form a large vector and the filter matrices are correspondingly resized 
    in a block-diagonal fashion.  
    For 3000 particles, this would be an 18000-element vector, 
    with corresponding matrices containing up to $18000\times18000$ elements.  
    At double precision, this amounts to approximately 2.5 gigabytes of storage per 
    time step, as well as many floating-point operations.  
    For this reason, we partitioned the myriad targets into subsets that were manageable 
    as individual multi-target trackers.

  \subsection{Interacting multiple model estimator}\label{sec:IMM} %
    The prediction step inherent in any Kalman-filter-based technique for dynamic state 
    estimation is based on a physical model for the underlying dynamics of the target, as 
    discussed for an EKF in the Section~\ref{sec:EKF}.  
    If the underlying dynamics change, then the filter performance (accuracy) will 
    decrease due to the reduced ability of the model to predict the changed/changing 
    dynamics of the target.  
    Such dynamical changes, called maneuvers, can often be described by adding an 
    input $u(k)$ to the prediction step of the filter.  
    Approaches to handling an unknown input fall into two broad categories\cite{BarShalom}:
    modeling the unknown input as a random process; 
    and estimating the unknown input in real time.  
    Input estimation can be computationally costly as it involves either maintaining a 
    history (sliding window) of estimates that is used to detect a maneuver, or it involves 
    increasing the state dimension by augmenting the state vector with the input to be estimated.  
    In the case where the underlying system dynamics is known to switch between a finite 
    number of modes, it can be significantly less costly to run a filter for each dynamical 
    model than to perform input estimation.  
    Such multiple model approaches to state estimation use a Bayesian framework where the prior 
    probabilities for each model being correct are updated to the posterior probability at each 
    point in time.  
    In an interacting multiple model\cite{Blom84,BlomBarShalom88} state estimator, the 
    mode probabilities $\mu_j$ are used in conjunction with tunable Markovian mode-switching 
    probabilities $p_{i,j}$ to calculate a mixed (weighted) initial condition at each time, 
    which is filtered using each possible current model.  
    All of this comes at a fraction of the cost of input estimation, with significantly 
    improved performance during maneuvers\cite{BarShalom} and the added benefit of being 
    decision-free (no maneuver detection is necessary).

    The IMM algorithm proceeds as follows.  
    Mixing probabilities $\mu_{i|j}$ are calculated from the probability $\mu_{i}(k-1)$ 
    that mode $i$ was in effect at time step $k-1$, 
    given that mode $j$ is in effect at time step $k$, 
    conditioned on the measurements up to $k-1$:
    \begin{equation}\label{eq:IMM_mixing_probabilities}
      \mu_{i| j}(k-1|k-1) = \frac{p_{ij} \mu_i(k-1)}{\bar{c}_j} \qquad i,j=1\ldots s,
    \end{equation}
    where $\bar{c}_j$ ensures normalization and we use $s=3$ modes (see below).  
    The mode probabilities $\mu_j(k)$ are stored along with the state vectors 
    $\hat{x}^j(k)$.  
    The mixing probabilities are used to calculate the mixed initial condition 
    \begin{equation}\label{eq:IMM_mixed_initial_condition}
      \hat{x}^{0j}(k-1|k-1) = \sum_{i=1}^s \hat{x}^i(k-1|k-1) \mu_{i|j}(k-1|k-1),
    \end{equation}
    which is input to the $j^\text{th}$ filter, along with the corresponding 
    covariance 
    \begin{eqnarray}
      P^{0j}
      &=& \sum_{i=1}^s \mu_{i|j}
        \left\{P^i + \sq{\hat{x}^i - \hat{x}^{0j}}
        \cdot \sq{\hat{x}^i - \hat{x}^{0j}}^T\right\} ,
        \label{eq:IMM_mixed_initial_condition_covariance}
    \end{eqnarray}
    where we have dropped the $(k-1|k-1)$ arguments for presentation purposes.  
    Each filter has a likelihood function associated with it, $\Lambda_j(k)$, which is 
    a Gaussian probability reflecting the likelihood of obtaining the new measurement $z(k)$, 
    given the predicted measurement using the mixed initial condition, 
    $\hat{z}^j(k|k-1) = H\cdot \hat{x}^{0j}(k-1|k-1)$.  
    After running the filters and calculating the likelihood functions, the mode 
    probabilities are updated to
    \begin{equation}\label{eq:IMM_mode_probability_update}
      \mu_j(k) = \frac{\Lambda_j(k) \bar{c}_j}{c} ,
    \end{equation}
    where $c$ is a normalization constant.
    
    We use a three-mode IMM, with a model each for before, during and after the shock wave.  
    The first model is an EKF based on our work in \Ref{Oxtoby2010:SPIE}, designed for 
    a dust crystal.  We refer to this as the base mode.  
    The second and third modes are referred to as the maneuvering modes, 
    where the target dynamics deviate from the base mode (usually for short periods of time).  
    The second mode includes an input $u(k)$ to account for an external force in the 
    positive-X direction, extending the prediction equation (\ref{nonlinStateTrans}) to 
    \begin{align}
      {\hat{x}}(k+1) &= f(\hat{x}(k_{+})) + G_\mathrm{y} u(k) \nonumber \\
        &= f(\hat{x}(k_{+})) 
          + \left( \begin{array}{c}(\Delta t)^2/2 \\ \Delta t \\ 1\\0\\0\\0 \end{array}\right)
            F_\mathrm{shock}(k) ,
      \label{eq:during}
    \end{align}
    where $G_\mathrm{y}$ is the input gain matrix for the unknown effective force of tunable 
    magnitude $u(k) = F_\mathrm{shock}(k)$.  The third mode is designed to account for particles 
    reflected in the opposite direction, and so the prediction equation has the same form as 
    (\ref{eq:during}), but with $u(k) = F_\mathrm{aftershock}(k)$ in the negative-X direction 
    (towards the left of the scene).  
    The remainder of the EKF equations are unchanged by the inclusion of these inputs.  
    For ephemeral maneuvers, the base mode should be the best description of the 
    dust dynamics for most of the time.  Thus, we choose the following mode-switching 
    probabilities 
    \begin{equation}
      p_{i,j} = \left(
        \begin{array}{ccc}
          0.80 & 0.10 & 0.10 \\
          0.30 & 0.60 & 0.10 \\
          0.40 & 0.10 & 0.50
        \end{array}
      \right) ,
    \end{equation}
    which sum to unity along each row, as required.  
    Larger values along the diagonal of $p_{i,j}$ lead to slower switching between modes.  
    This can also be tuned in the opposite direction if faster mode switching is desired.  
    Faster mode switching tends to yield lower peak RMS error during a maneuver, at the 
    expense of greater RMS error during quiescent periods.\cite{BarShalom}  
    Initially, the mode probabilities are equal to one other: $\mu_1=\mu_2=\mu_3=1/3$.  
    IMM mode probabilities can be used to visualize the maneuvers, as done for a 
    shocked dusty plasma in \Ref{Oxtoby2011:IEEE:a}.

\subsection{Range of influence}\label{sec:NN}%
  The Yukawa force in \Eq{eq:force} decreases both exponentially and polynomially with 
  increasing particle separation, thereby reducing the range of influence that each dust 
  particle has on other particles.  
  Beyond this range, two dust particles exert negligible forces on each other.  
  Here we define the range of influence $\tilde{r}_\mathrm{max}$ relative to the nearest 
  neighbor separation $\tilde{r}_\mathrm{NN}$ as the separation at which the two-particle 
  Yukawa force magnitude drops to $1\%$ of the Yukawa force due to the nearest neighbor: 
  $F_{\jmath,k}(\tilde{r}_\mathrm{max}) / F_{\jmath,k}(\tilde{r}_\mathrm{NN}) = 1\%$.  
  This occurs for $\tilde{r}_\mathrm{max} \approx 3.81 \tilde{r}_\mathrm{NN}$.  
  This range of influence concept was used to simplify the particle tracking.  
  The cutoff was not used when numerically integrating the equations of motion.


\end{document}